\documentclass[aps,pra,twocolumn,groupedaddress,showpacs,superscriptaddress,amssymb,amsmath]{revtex4-2}
\usepackage{pgfplots}
\pgfplotsset{compat=1.13}
\usepackage[utf8]{inputenc}
\usepackage{graphicx}
\usepackage{tabularx}
\usepackage{xcolor}
\usepackage{amsmath}

\usepackage{comment}
\usepackage{dcolumn}
\usepackage{hyperref}
\hypersetup{
    colorlinks=true,
    linkcolor=blue,
	citecolor=cyan
}
\usepackage{bm}
\usepackage{epsf}
\usepackage{braket}
\usepackage{tensor}
\usepackage{soul}
\usepackage{float}
\usepackage{subfig}
\usepackage{mathrsfs}
\usepackage{mathtools}
\usepackage{multirow}
\usepackage{caption}
\usepackage{amsfonts}

\newcommand{\be}{\begin{equation}}
\newcommand{\ee}{\end{equation}}
\newcommand{\bea}{\begin{eqnarray}}
\newcommand{\eea}{\end{eqnarray}}

\makeatletter
\newsavebox{\@brx}
\newcommand{\llangle}[1][]{\savebox{\@brx}{\(\m@th{#1\langle}\)}%
  \mathopen{\copy\@brx\kern-0.5\wd\@brx\usebox{\@brx}}}
\newcommand{\rrangle}[1][]{\savebox{\@brx}{\(\m@th{#1\rangle}\)}%
  \mathclose{\copy\@brx\kern-0.5\wd\@brx\usebox{\@brx}}}
\makeatother

\begin{document}
\title{Modified quantum regression theorem and consistency with Kubo-Martin-Schwinger condition}
\author{Sakil Khan}
\email{sakil.khan@students.iiserpune.ac.in}
\affiliation{Department of Physics,
		Indian Institute of Science Education and Research, Pune 411008, India}
\author{Bijay Kumar Agarwalla}
\email{bijay@iiserpune.ac.in}
\affiliation{Department of Physics,
		Indian Institute of Science Education and Research, Pune 411008, India}
\author{Sachin Jain}
\email{sachin.jain@iiserpune.ac.in}
\affiliation{Department of Physics,
		Indian Institute of Science Education and Research, Pune 411008, India}
\date{\today}
\begin{abstract}
We show that the long-time limit of the two-point correlation function obtained via the standard quantum regression theorem, a standard tool to compute correlation functions in open quantum systems, does not respect the Kubo-Martin-Schwinger equilibrium condition to the non-zero order of the system-bath coupling. We then follow the recently developed Heisenberg operator method for open quantum systems and by applying a ``{\it weak}" Markov approximation, derive a new modified version of the quantum regression theorem that not only respects the KMS condition but further predicts exact answers for certain paradigmatic models in specific limits. We also show that in cases where the modified quantum regression theorem does not match with exact answers, it always performs better than the standard quantum regression theorem. 
\end{abstract}
  \maketitle  


\section{Introduction}
Understanding the quantum dynamics and  subsequent long-time state of a system coupled to a finite-temperature environment is an active area of research \cite{xiong2020exact,PhysRevB.107.125149,PhysRevE.100.022111,PhysRevB.97.134301,PhysRevB.86.155424}. Often, except for a few particular cases, it is difficult to analyze such open quantum systems exactly. One therefore resorts to various approximate descriptions. One such example is the well-known quantum master equation (QME) for the reduced density matrix of a system which is typically derived under the weak system-environment coupling  assumption and further making the Markovian approximation \cite{Carmichael,breuer,Alicki,davies,REDFIELD19651,rivas2012open,lidar2019lecture}.  If one subsequently applies yet another approximation (called the secular approximation) one obtains the popular Lindblad master equation \cite{GORINI1978149,Lindblad,Gorini}.
The QME approach has been extensively applied to understand open quantum dynamics and quantum transport properties \cite{wald2018lindblad,PhysRevE.94.032139,esposito2007quantum,PhysRevB.74.235309,PhysRevE.68.066112,RevModPhys.93.015008,RevModPhys.70.101}. In recent years, efforts have been put forward to analyze systems beyond the approximate regimes \cite{PhysRevE.105.034112,PRXQuantum.4.020307,PhysRevA.93.062114,PhysRevB.92.235440,PhysRevLett.114.080602,PhysRevA.104.052617,Anto-Sztrikacs_2021,PhysRevA.99.010102,PhysRevB.84.161414,PRXQuantum.3.010321,PhysRevE.101.052129,PhysRevLett.112.110401,PhysRevX.10.041024,trushechkin2022quantum,PhysRevA.101.012103,PhysRevA.90.032114,PhysRevB.83.115416} and also to perform consistency checks against well-known universal symmetries such as fluctuation symmetries \cite{fluc-1, fluc-2,Jar-fluc,segal-1,GKLS-latest,agarwalla2016reconciling}.

Typically, systems once reach thermal equilibrium in the long-time limit are supposed to follow certain universal relations such as the fundamental Kubo-Martin-Schwinger (KMS) equilibrium condition for two or multi-time correlation functions \cite{kubo1957statistical, MS, breuer,Carmichael}. 
For open quantum systems, one of the most powerful tools to compute such multi-time correlation functions is the well-known Quantum Regression Theorem (QRT) \cite{Carmichael,breuer,qn,otoc1,qrt,Sakil1}.  The validity of the QRT crucially relies on weak system-bath coupling and the  Markovian approximation \cite{Carmichael,breuer}. 
More precisely, to derive QRT for the two-point correlation function of the form $\big\langle O_{1}(t+\tau)O_{2}(t)\big\rangle$, one needs to assume $\tau \gg \tau_{B}$ where $\tau_B$ is the characteristic time scale of the bath \cite{Carmichael, breuer}. This particular condition from here onwards we will refer to as the ``{\it strong}'' Markovian approximation. 
However, we show that the correlation function obtained using the standard QRT (SQRT) fails to preserve the KMS equilibrium condition at the long-time limit to the leading order in system-bath coupling. It is therefore important to ask, is there a modified version of the SQRT whose long-time solution respects the KMS condition to the non-zero order of the system-bath coupling?

In this work, we provide a detailed answer to this question. By relaxing the strong Markovian approximation, we obtain a modified version of QRT (MQRT) that perfectly respects the KMS condition in equilibrium. In other words, we impose a {\it weaker} Markov condition for computing the correlator $\langle O_{1}(t+\tau)O_{2}(t)\rangle$, given as $t+\tau \gg\tau_{B}$ instead of $\tau \gg \tau_{B}$. Moreover, we show that for certain paradigmatic models, interestingly, the MQRT predicts exact answers under specific limits. We further show that in cases when the MQRT does not predict the exact answer,  it always works better than the SQRT. 
 
We organize the paper as follows: In section \ref{Sec-II}, we provide a brief background on the SQRT and discuss the consistency of KMS condition in the long-time limit. We show that following the SQRT, the KMS condition is violated to the finite order of the system-bath coupling. In section \ref{Sec-III}, we give the details of the modified version of the QRT. In section \ref{Sec-IV} we provide a model example to illustrate the MQRT and explicitly show how the KMS condition is preserved in equilibrium. We further show that the MQRT predicts exact answers in specific limits of the model and more importantly in all cases works better than the SQRT. We then summarize our work in section \ref{Summary}. We delegate certain details to the appendices, such as the detailed derivation of the MQRT, another model example to illustrate further the MQRT and consistency with the KMS, as well as the extension of MQRT to three-point correlators.


\section{A brief background on SQRT and consistency check in equilibrium}
\label{Sec-II}
 We start by writing down the Hamiltonian of the total system as $H  = H_{S}+H_{R}+ H_{S  R}$, where $H_{S}$ ($H_{R}$) represents the bare system (reservoir)  Hamiltonian and $H_{S R}$ represents the  system-reservoir interaction Hamiltonian. We assume that initially at $t=0$, the system and the reservoir were decoupled and therefore the total density matrix can be written as $\rho_{SR}=\rho_{S} \otimes \rho_{R}$.  In the  Heisenberg picture, 
 the expectation value of any system operator can be written as  $\langle O(t)\rangle={\rm Tr}_{S}[{\rm Tr}_{R}[O(t)\rho_{R}]\rho_{S}]={\rm Tr}_{S}[O_{S}(t)\rho_{S}]$, where we define the reduced one point system operator as $O_{S}(t)= {\rm Tr}_{R}[O(t)\rho_{R}]$.
The dynamics of such a reduced one-point system operator is governed by the following adjoint quantum master equation \cite{Carmichael,breuer,karve2020heisenberg} 
\begin{align}\label{ad-QME}
	&\frac{d}{dt}O_{S}(t)=i \Big[H_{S}+H_{LS}, O_{S}(t)\Big] \nonumber\\
	&+ \sum_{k} \gamma_{k} \Big(  L^{\dagger}_{k} O_{S}(t) L_{k}-\frac{1}{2}\Big\{L^{\dagger}_{k} L_{k}, O_{S}(t)\Big\} \Big)\;,
\end{align} 
where $L_{k}$'s are known as the quantum jump operators and $\gamma_{k}$'s are positive number i.e.,  $\gamma_{k}\geq 0$. Note that, the adjoint master equation  in Eq.~\ref{ad-QME} is derived under the following assumptions: (a) the interaction between the system and the bath is assumed to be weak (Born approximation), and the effect upto only up to the second order of $H_{S R}$ is incorporated in the dynamics, (b) Markovian approximation i.e., one assumes $t\gg\tau_{B}$, where $\tau_{B}$ is the  characteristic time scale of the bath correlation function, and (c) the Secular approximation \cite{Carmichael,breuer,karve2020heisenberg}.

Using the adjoint master equation in Eq.~\ref{ad-QME}, one can compute the expectation value of any system operator. However, in computing the correlation functions, QRT turns out to be the most useful tool \cite{breuer,Carmichael, davies,Alicki,qn}.
Briefly, the QRT states that the knowledge of the time evolution of a single-point function is sufficient to determine the time evolution of two-point or multi-point  correlation functions. More explicitly, let's assume that there exists a complete set of system operators $A_{\mu}, \mu=1,2,...$ such that 
\begin{equation}\label{coma}
	\frac{d}{dt} \langle A_{\mu}(t)\rangle =\sum_{\kappa} M_{\mu \kappa}\langle A_{\kappa}(t)\rangle.
\end{equation}
Then the standard QRT (SQRT) for the two-point function reads as 
\begin{equation}
\label{SQRT}
\frac{d}{d\tau}\langle A_{\mu}(t+\tau) \, O(t)\rangle =
	\sum_{\kappa} M_{\mu \kappa}\langle A_{\kappa}(t+\tau) O(t)\rangle.
\end{equation}
The above SQRT can be systematically extended for $N$-point function \cite{Sakil1}.
Recall that to derive the SQRT in Eq.~\ref{SQRT}, apart from assuming weak coupling between the system and bath, the ``{\it strong}'' Markovian limit i.e., $\tau \gg \tau_B$ is assumed. 
One of the consistency criteria that the correlation function must satisfy in thermal equilibrium (i.e., in the limit $t \to \infty$) is the KMS condition which for two-point correlators of arbitrary system operators is given as 
 \begin{equation}
 \label{KMS}
 	\langle O_{1}(\tau)O_{2}(0)\rangle_{\rm eq}=\langle O_{2}(0)O_{1}(\tau+i\beta)\rangle_{\rm eq}\;,
 \end{equation}
where $\beta=\frac{1}{k_B T}$ is the inverse temperature of the bath. It can be shown that the solution of the correlation function obtained following SQRT in Eq.~\ref{SQRT} satisfies the KMS condition up to the zeroth order, and fails to satisfy the KMS condition at any finite order of the system bath coupling. 
Let us illustrate this point with a simple example. We consider a simple toy model consisting of a single bosonic degree of freedom of frequency $\omega_0$ which is further coupled to a bosonic bath of inverse temperature $\beta$, as given in Eq.~\ref{CL}. One can easily compute the following two-point correlation function using the SQRT in Eq.~\ref{SQRT}, given as \cite{Carmichael},
\begin{eqnarray}
	&	\langle a^\dagger(t+\tau)a(t)\rangle =	\langle a^\dagger(t)a(t)\rangle \,\, e^{(i\omega'_{0}-\frac{\gamma}{2})\tau},\nonumber\;\\
		&\langle a(t)a^\dagger(t+\tau)\rangle=	\langle a(t)a^\dagger(t)\rangle \,\, e^{(i\omega'_{0}-\frac{\gamma}{2})\tau},
  \label{two-example}
\end{eqnarray}
where $\gamma$ is related to the system-bath coupling and $\omega_0'$ is the renormalized frequency of the system due to interaction with the bosonic bath. In the equilibrium limit i.e., in the limit $t \to \infty$, we receive from Eq.~\ref{two-example}, 
\begin{align}
&	\lim_{t \to \infty} \langle a^\dagger(t+\tau)a(t)\rangle=	\langle a^\dagger a\rangle_{(\infty)} \,e^{(i\omega'_{0}-\frac{\gamma}{2})\tau}\;,\nonumber\\
&\lim_{t \to \infty} 	\langle a(t)a^\dagger(t+\tau)\rangle=	\langle a a^\dagger\rangle_{(\infty)} \,e^{(i\omega'_{0}-\frac{\gamma}{2})\tau},
\label{long-correlator}
\end{align}
which reduces to a function of only the time difference $\tau$, as expected in equilibrium. Now, it is clear from Eq.~\ref{long-correlator} that the correlators satisfy the KMS condition (Eq.~\ref{KMS}) only when $\gamma$ is set to zero i.e., only upto zeroth order of system-bath coupling and fails to respect KMS at any finite-order of the system-bath coupling. Note that to validate the KMS we use the fact that in equilibrium, at the zeroth order of system-bath coupling, $\langle a a^{\dagger}\rangle_{\rm eq} = e^{\beta \omega_0} \langle a^{\dagger} a \rangle_{\rm eq}$.
So a natural question that immediately arises from the above exercise is how to respect the KMS condition in equilibrium that includes finite corrections due to system-bath coupling. In order to achieve this, in what follows we put forward a modified version of the QRT,  which from here onward we refer to as MQRT, by incorporating a {\it weaker} Markovian approximation.

\section{MQRT under weak Markov approximation}
\label{Sec-III}
In this section, we provide the details about the MQRT, and the main steps of the derivations are provided in Appendix \ref{app-1}. To proceed further, We consider a generic form of the total Hamiltonian as,
\begin{equation} \label{eq1nm}
			H = H_{S}+\sum_{k} \Omega_{k} b_{k}^\dagger b_{k}+\lambda \sum_{k} \alpha_{k}( S \, b_{k}^\dagger+ S^\dagger  b_{k})\;,
\end{equation}
where the second term in Eq.~\ref{eq1nm} represents the bath Hamiltonian consisting of infinite number of non-interacting oscillators with $b_{k} $ and $b_{k}^\dagger $ can in general represent  the bosonic or fermionic annihilation and creation operator for the $k$-th mode, respectively. The corresponding frequency of the $k$-th mode is denoted by $\Omega_k$. The third term represents the system-bath coupling with the generic system operator $S$ coupled with the $k$-th bath mode with interaction strength $\alpha_k$. We introduce the parameter $\lambda$ to keep track of the order of the perturbation explicitly.  

We first start by assuming that the expectation value of some arbitrary system operators satisfies Eq.~\ref{coma}.
Our motivation is to calculate the two-point correlation function of the form
\begin{equation}
\langle A_{\mu}(t+\tau)O(t) \rangle={\rm Tr}_{S}\Big[[A_{\mu  }(t+\tau)O(t)]_{S} \,\rho_{S}(0)\Big],
\end{equation}
where $[A_{\mu  }(t+\tau)O(t)]_{S}$
denotes the two-point reduced operator \cite{Sakil1,karve2020heisenberg}. We follow the prescription of Ref.~\cite{karve2020heisenberg}, to express the two-point reduced operator $[A_{\mu  }(t+\tau)O(t)]_{S}$, up to order $\lambda^2$, in terms of one-point reduced operator and an irreducible part (for details see Appendix-\ref{app-1}).

We can show that under the {\it weak} Markov approximation i.e., $t+\tau \gg \tau_{B}$ \cite{PhysRevA.105.062204}, where $\tau_{B}$ is the bath characteristic time scale, the two-point reduced operator  $ [A_{\mu  }(t+\tau)O(t)]_{S}$ satisfies the following equation 
\begin{widetext}
	\begin{align} \label{eqqrtcc}
			\frac{d}{d\tau}\Big[A_{\mu  }(t+\tau)O(t)\Big]_{S} \!=&\sum_{\kappa}M_{\mu\kappa}\, \Big[A_{\kappa }(t+\tau)O(t)\Big]_{S} +C_{1}(t, \tau)+C_{2}(t, \tau).
	\end{align}
 We provide the details of this derivation in Appendix-\ref{app-1}. Note that, to derive the above equation we have considered the Born approximation i.e., we collect terms up to the $\lambda^2$ order and the weak Markov condition. The correction to the SQRT shows up by two new non-homogeneous  terms $C_{1}(t, \tau)$  and  $C_{2}(t, \tau)$ which are given by the following equations
\begin{align}
\label{C1}
	&C_{1}(t, \tau) \!=\! \lambda^2 \!\! \sum_{j,l,m,n} \!\! e^{-i(\omega_{j}+\omega'_{m})(t+\tau)} e^{-i(\tilde{\omega}_{l}-\omega'_{n})t}\int_{0}^{\infty}\! d\tau'_{2}\;  e^{i(\tilde{\omega}_{l}-\omega'_{n})\tau'_{2}}  \Big[ A_{\mu j S}(0),
	S_{m}\Big]\Big [ S^\dagger_{n}, O_{l S}(\tau'_{2})\Big] \int_{-\infty}^{\infty} \frac{ d\Omega}{2\pi}  F_{\eta}(\Omega)\;e^{i\Omega (\tau+\tau'_{2})}\\
 &	C_{2}(t, \tau)\!=\! \lambda^2 \!\!\sum_{j,l,m,n} \!\!  e^{-i(\omega_{j}-\omega'_{m})(t+\tau)} e^{-i(\tilde{\omega}_{l}+\omega'_{n})t}\int_{0}^{\infty}\!  \!d\tau'_{2} e^{i(\tilde{\omega}_{l}+\omega'_{n})\tau'_{2}}  \! \Big[ A_{\mu j S}(0),
	S^\dagger_{m}\Big]\Big [ S_{n}, O_{l S}(\tau'_{2})\Big]  \!\! \int_{-\infty}^{\infty} \! \frac{ d\Omega}{2\pi} \Big(J(\Omega) -  \eta \,   F_{\eta}(\Omega)\Big) e^{-i\Omega (\tau+\tau'_{2})}
 \label{C2}
\end{align}
\end{widetext}
where $J(\Omega)$ is the spectral density function of the bath which is defined as $J(\Omega)= 2\pi \sum_{k} |\alpha_{k}|^2 \delta(\Omega-\Omega_{k})$ and $F_{\eta}(\Omega)=J(\Omega) n_{\eta}(\Omega)$ with $  n_{\eta}(\Omega)=[ e^{\beta \Omega}+ \eta]^{-1}$ with $\eta=+1$ and $\eta=-1$ are for fermions and bosons, respectively. $\omega_j$ corresponds to the possible energy differences between the bare system eigenenergies that appear by performing spectral decomposition for the operator $A_{\mu S}$. In other words, we use the fact that 
\begin{equation}\label{specttral1}
A_{\mu S}(t)=\sum_{j}  A_{\mu j  S}(t-t') e^{-i\omega_{j}t'}+O(\lambda^2).
\end{equation}
Similarly, $\tilde{\omega}_l$ and $\omega_m'$ correspond to the possible energy differences of the bare system for the operators $O_{S}$ and $S$ respectively. Eq.~\ref{eqqrtcc} is one of the central results of this paper.

Note that MQRT in Eq.~\ref{eqqrtcc} reduces to SQRT once  the stronger Markovian approximation $\tau \gg \tau_{B}$ is incorporated. This can be seen as follows. The integration over $d\Omega$ in Eqs.~\ref{C1} and \ref{C2} involves the bath correlation function.  Once we assume  $\tau \gg \tau_{B}$ then it is clear that for every value of $\tau'_{2}$ appearing in the integrand as $e^{-i\Omega (\tau+\tau'_{2})}$
the bath correlation function or the $d\Omega$ integration vanishes since $\tau +\tau'_{2} \gg \tau_{B}$ where $\tau_B$ is the bath correlation decay time. As a result both $C_{1}(t, \tau)$  and  $C_{2}(t, \tau)$ vanishes and one recovers the SQRT. 

Our analysis presented here can be systematically extended to obtain MQRT for higher-order correlators as well. In Appendix \ref{app-3}, we discuss a model example for MQRT that involves three point correlators.  In what follows we will show that the correlation functions obtained from the MQRT satisfies the KMS condition in the long-time limit. Interestingly, we also find that in specific limits, the above derived MQRT matches with exact answers. Also, for situations when the MQRT deviates from exact answer it still makes better predication than the SQRT in Eq.~\ref{SQRT}.
 \section{Example}
 \label{Sec-IV}
In this section, we  apply the MQRT given in Eq.~\ref{eqqrtcc} to compute the correlation function for a dissipative non-interacting bosonic/fermionic system. We further illustrate the use of Eq.~\ref{eqqrtcc} for the dissipative spin-boson model in Appendix \ref{app-2}. For both these models, we  show explicitly that the correlation functions obtained using MQRT satisfy the KMS condition in the long-time limit whereas the SQRT fails drastically. Moreover, at equilibrium, the correlation functions received following MQRT are closer to the exact result than the QRT.

\subsection{Dissipative Bosonic/Fermionic Model}
We consider a single bosonic or fermionic degree of freedom as a system that is coupled to a corresponding bosonic or fermionic thermal bath. The total Hamiltonian is given by
\begin{equation} \label{CL}
		H
		= \omega_{0} a^\dagger a+\sum_{k} \Omega_{k} b_{k}^\dagger b_{k}+ \sum_{k} \alpha_{k}( a\, b_{k}^\dagger+a^\dagger \, b_{k}),
\end{equation}
where $a $ and $a^\dagger $ represent  the bosonic or fermionic annihilation and creation operator for the system, respectively. The second term represents the environment and the third term represents system-environment coupling. All the operators appear in Eq.~\ref{CL} follow either the bosonic commutation or fermionic anti-commutation algebra. 
For this setup, using the adjoint Lindblad QME in Eq.~\eqref{ad-QME}, one obtains for the reduced one-point operators $a_{S}(t)$ and  $a^{\dagger}_{S}(t)$ the following equations
\begin{align}
\frac{d}{dt}\begin{bmatrix} a_{S}(t)  \\ a^{\dagger}_{S}(t)\end{bmatrix}=\begin{bmatrix}  -G^{*}  & 0 \\ 0& -G\end{bmatrix}
\begin{bmatrix}  a_{S}(t)  \\ a^{\dagger}_{S}(t)\end{bmatrix},
\end{align}
where $G=(-i\,\omega'_{0}+\frac{\gamma}{2})$ with $\omega_0'=\omega_{0}+P \big[\int \frac{d\Omega }{2 \pi}\frac{J(\Omega)}{\omega_{0}-\Omega}\big]$  is the modified frequency where $P$ denotes the Cauchy principle value and $\gamma=J(\omega_{0})$ where recall that $J(\omega)$ is the spectral density of the bath.  We first compute the  correlation function, $\langle a^\dagger(t+\tau)a(t) \rangle$ using the MQRT in \eqref{eqqrtcc}.
In the MQRT, since $C_{1}(t,\tau)$ and $C_{2}(t,\tau)$ are the additional terms, we explicitly show how to  compute these terms. Let us first note that for this example, spectral decomposition in \eqref{specttral1} takes a very simple form and it has only one characteristic frequency.
Comparing with Eq.~\ref{eqqrtcc}-\ref{C2} we note, $A_{\mu }=a^\dagger$, $O=a$. With this identification it is easy to show $A_{\mu 1 S}(0)=a^\dagger$, $O_{1 S}(0)=a$ with $\omega_{1}=-\omega_{0}$
and  $\tilde{\omega}_{1}=\omega_{0}$.
Similarly for this model, 
$S=a$ and  it is easy to show that $S_{1}=a$ with $\omega'_{1}=\omega_{0}$.
Substituting these quantities in Eq.\eqref{C1}, we obtain the first non-homogeneous part of Eq. \eqref{eqqrtcc} as
\begin{equation}
	C_{1}(\tau)\!=\![a^\dagger,a ][a^\dagger,a ]\!\int_{-\infty}^{\infty} \frac{ d\Omega}{2\pi}  F_{\eta}(\Omega)\, e^{i\Omega \tau} \! \int_{0}^{\infty} d\tau'_{2}  e^{(i\Omega-G^{*}) \tau'_{2}}
\end{equation}
which reduces to the following equation after performing the $\tau'_{2}$ integration,
\begin{equation}
	C_{1}(\tau)=  \int_{-\infty}^{\infty} \frac{ d\Omega}{2\pi}  \;e^{i\Omega \tau} \frac{  F_{\eta}(\Omega)}{G^{*}-i\Omega}.
\end{equation}
Recall that the subscript of $\eta$ in $F_{\eta}$ indicates Fermi or Bose statistics. Note that interestingly in this case, $C_{1}(t,\tau)$ depends only on $\tau$ and all $t$ dependence in Eq.\eqref{C1}  disappears. The other non-homogeneous term i.e., $C_{2}(t,\tau)$ in Eq.~\eqref{C2} for this setup vanishes as,
 \begin{equation}
     C_{2}(t,\tau) \propto [a^\dagger,a^\dagger ][a,a ]=0.
 \end{equation}
Finally, the MQRT for the two-point reduced operator $\langle a^\dagger(t+\tau)a(t) \rangle$ takes the following form
\begin{equation} 
		\frac{d}{d\tau}\big[a^{\dagger}(t+\tau) a(t)\big]_{S} \!=-G\,\big[a^{\dagger}(t+\tau) a(t)\big]_{S}+C_{1}(\tau).
\end{equation}
Let us note that the MQRT contains one crucial non-homogeneous term $C_1(\tau)$ in this case, which is missed in the SQRT, see \eqref{coma}. This extra term is the direct consequence of the {\it weak} Markovian approximation. The formal solution of the above equation is given by
\begin{align} 
		\big[a^{\dagger}(t+\tau) a(t)\big]_{S} \!=& e^{-G \tau}\,\big[a^{\dagger}(t) a(t)\big]_{S}\nonumber\\
		&+\int_{-\infty}^{\infty} \frac{ d\Omega}{2\pi}  \; \frac{(e^{i\Omega \tau} -e^{-G \tau}) F_{\eta}(\Omega)}{(G^{*}-i\Omega) (G+i\Omega)}.
\end{align}
We can then calculate $ \langle a^\dagger(t+\tau)a(t) \rangle$ at the equilibrium by taking $t\to \infty$ limit. We obtain,  
\begin{equation} \label{eqada}
		\langle a^\dagger(t+\tau)a(t) \rangle^{\text{MQRT}}_{\rm eq} \!=\int_{-\infty}^{\infty} \frac{ d\Omega}{2\pi}  \; \frac{e^{i\Omega \tau} n_{\eta}(\Omega) J(\Omega)}{(G^{*}-i\Omega) (G+i\Omega)}.
\end{equation}
Note that, to get the above expression we need to first calculate $ \langle a^\dagger(t)a(t) \rangle$ at equilibrium, 
the calculation of which is given in Appendix \ref{app-4}. Similarly, one can compute the other correlator,	$\langle a(t) a^\dagger(t+\tau) \rangle$, which is given by
\textcolor{black}{\begin{align} \label{eqaad}
		\langle a(t) a^\dagger(t+\tau) \rangle^{\text{MQRT}}_{\rm eq} \!=\int_{-\infty}^{\infty} \frac{ d\Omega}{2\pi}  \; \frac{e^{i\Omega \tau} \big (1 - \eta \, n_{\eta}(\Omega)\big)J(\Omega)}{(G^{*}-i\Omega) (G+i\Omega)}\;,
\end{align}}
where recall 
From Eq.\eqref{eqada} and Eq.\eqref{eqaad}, it is clear that 
the above correlators satisfy the KMS condition up to the second order in system-reservoir coupling at equilibrium.

Following exactly the same procedure, one can also compute the correlator $\langle N (t+\tau) N(t)\rangle$ where $N(t)=a^{\dagger}(t) a(t)$ is the number operator. In the long-time limit, we receive,
\begin{align} \label{n-n}
	& \langle N (t+\tau) N(t)\rangle^{\text{MQRT}}_{\rm eq} \!=\int_{-\infty}^{\infty} \frac{ d\Omega}{2\pi}  \; \frac{J(\Omega)}{(G^{*}-i\Omega) (G+i\Omega)}\nonumber\\
	&\quad \quad \Big[n_{\eta}(\omega_{0}) n_{\eta}(\Omega)+ n_{\eta}(\omega_{0}) \big(1-\eta \,n_{\eta}(\Omega)\big) e^{i(\omega_{0}-\Omega)\tau} \nonumber\\
	& \quad \quad + n_{\eta}(\Omega) \big(1 - \eta \, n_{\eta}(\omega_{0})\big)\,e^{-i(\omega_{0}-\Omega) \tau}  \Big]\;
\end{align}
which once again
satisfies the KMS condition exactly. Note that, if one follows the SQRT in Eq.~\ref{SQRT}, the above correlation function gives,
\begin{align}
 \langle N(t+\tau) N(t)\rangle^{\text{SQRT}}_{\rm eq} \!=
	\big[\langle N  N \rangle_{(\infty)} -R \big]\, e^{-\gamma\tau} + R,
\end{align}
where $R=\int_{-\infty}^{\infty} \frac{ d\Omega}{2\pi}  \; \frac{F_{\eta}(\Omega) n_{\eta}(\omega_{0})}{(G^{*}-i\Omega) (G+i\Omega)}$. Clearly, this expression does not satisfy the required KMS to the non-zero order of the coupling (finite $\gamma$).
\begin{figure}
        \hspace*{0.47cm}
        \includegraphics[scale=.675]{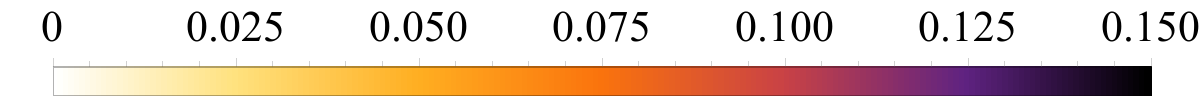}
        \\
        \vspace*{-0.4cm}
	\subfloat[]{\includegraphics[scale=.35]{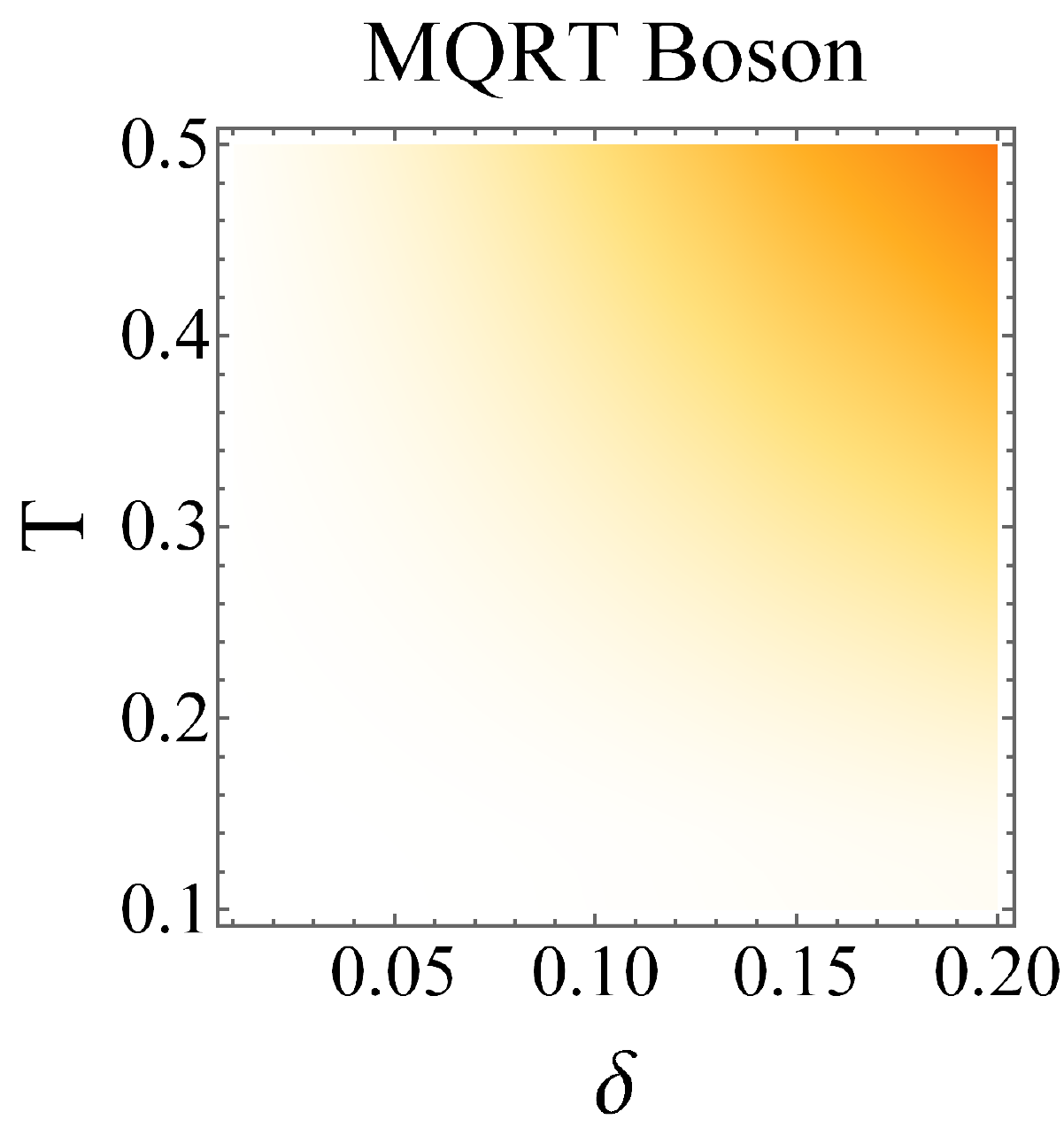}}
	\subfloat[]{\includegraphics[scale=.35]{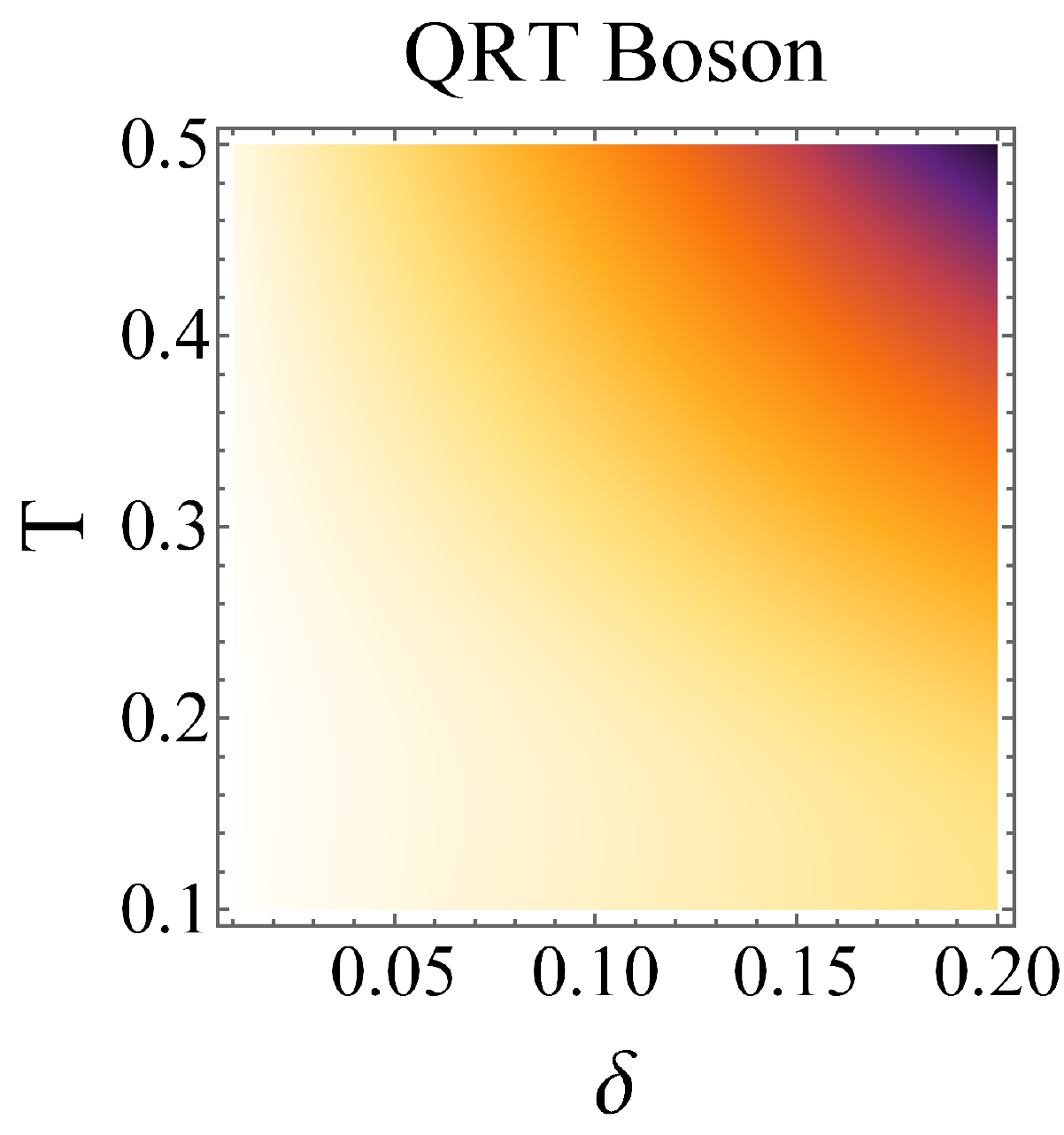}}
\\
\vspace*{0.4cm}
        \hspace*{0.5cm}
        \includegraphics[scale=.66]{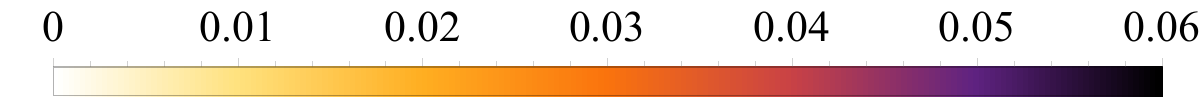}
        \\
        \vspace*{-0.4cm}
	\subfloat[]{\includegraphics[scale=.35]{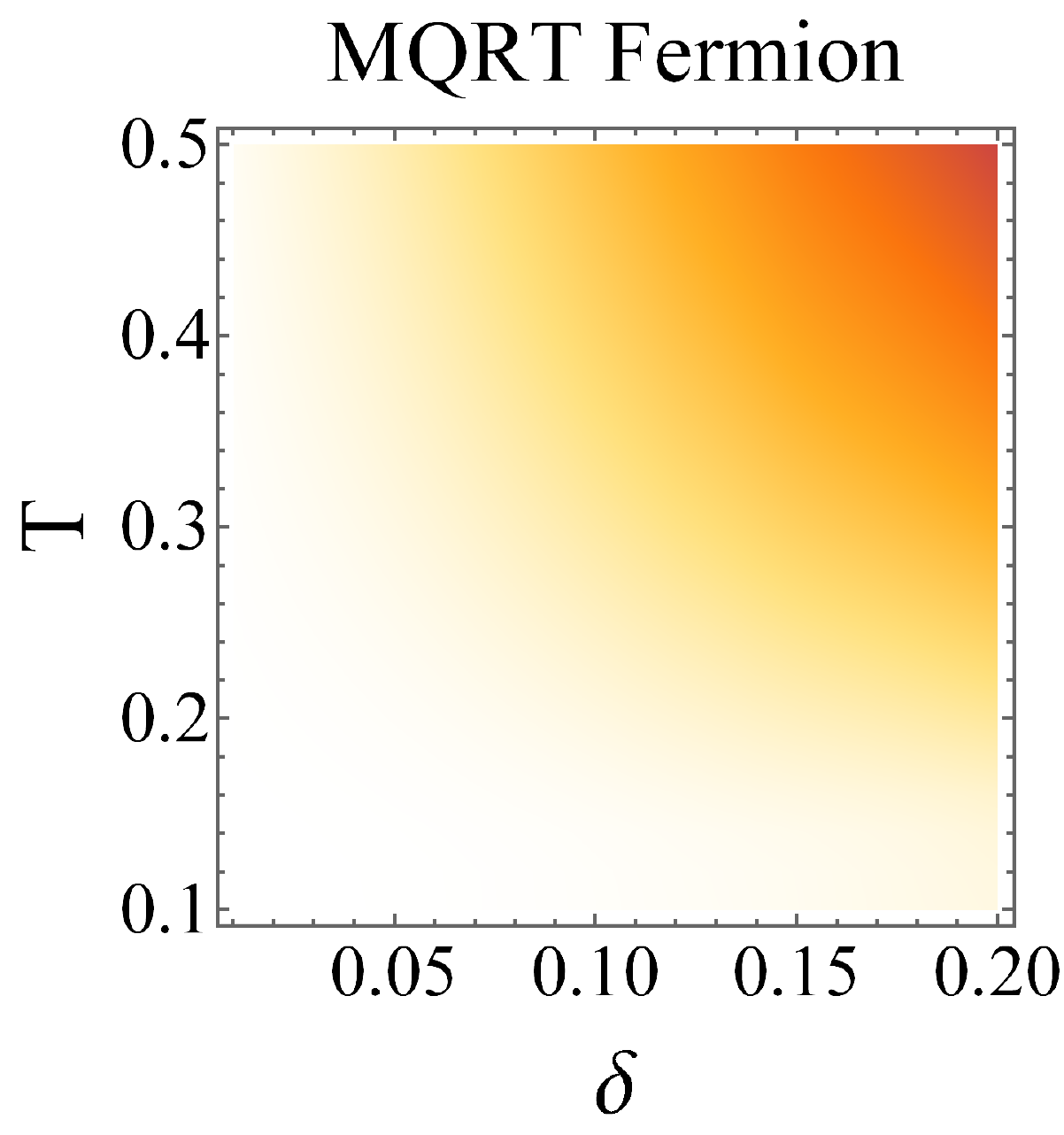}}
		\subfloat[]{\includegraphics[scale=.35]{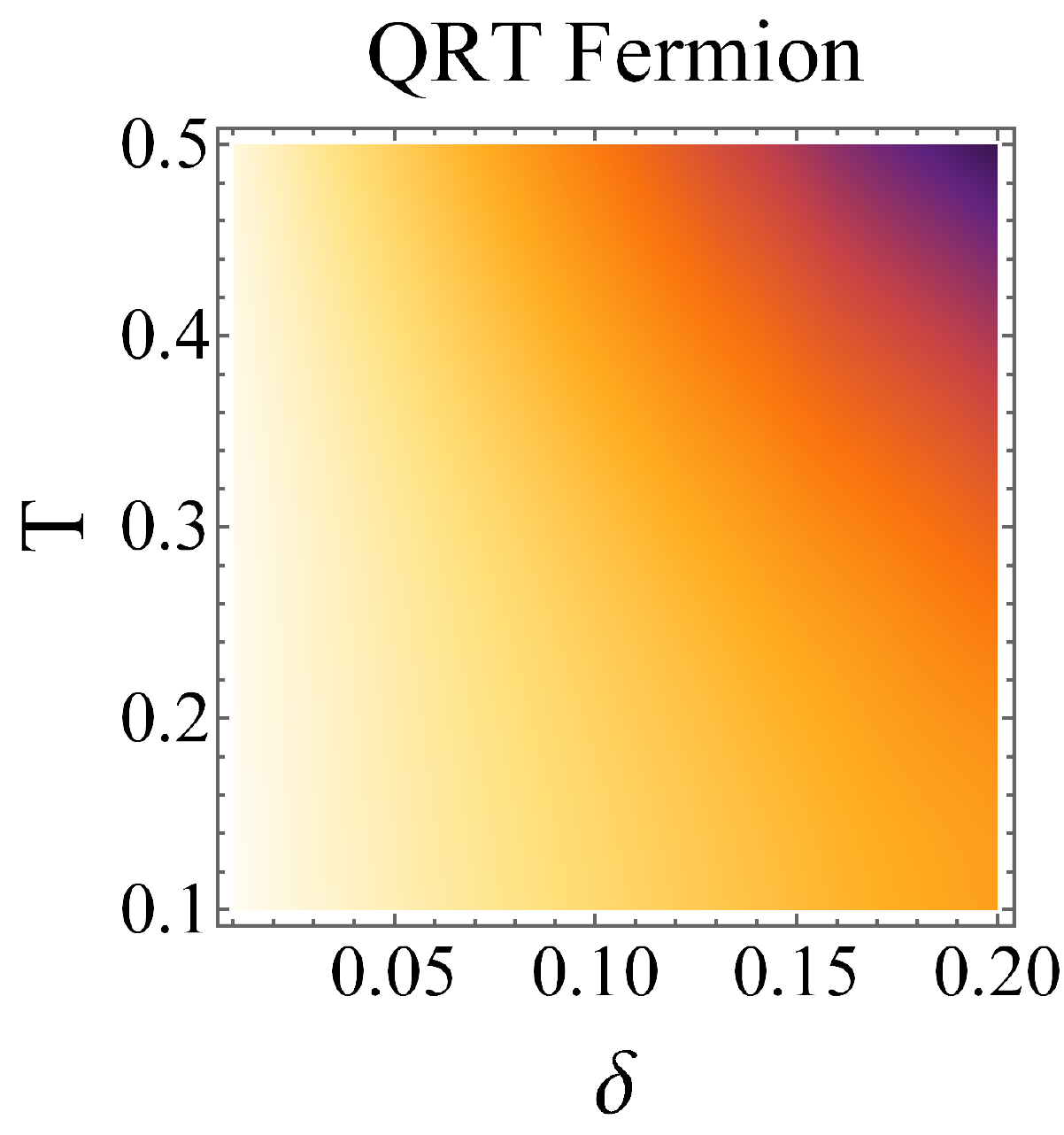}}
    \caption{(Color online:) Density plot for the quantity $D$, defined in Eq.~\ref{diff}, with respect to the temperature $T$ and $\delta$ defined via the spectral density $J(\Omega)$ taken as  $J(\Omega)= 2 \pi \delta \frac{\Omega^2}{1+\Omega^4}$. Results for the dissipative boson   model which captures the deviation of  the two-point  correlation function from the exact answer for (a) MQRT and (b) SQRT. Similarly, (c) and (d) are the results of the dissipative fermion model. We fix $\gamma=\pi$, $\omega_0=1$ and the upper integration limit $\tau_f$ in Eq.~\ref{diff} as $1/\delta$. In
    both cases, we observe that the answers predicted by MQRT are always closer to the exact result than the SQRT.}
	\label{fig:0}
\end{figure}

As the model in Eq.~\ref{CL} is exactly solvable, we now discuss the result obtained in equilibrium following the MQRT with the exact answer. Following the quantum Langevin equation approach \cite{QLE-ford} or the non-equilibrium Green's function technique \cite{ADDS,wang2014nonequilibrium}, one can readily obtain the exact long-time result for the correlation function as, 
\begin{equation} \label{exact1}
\langle a^\dagger(t+\tau)a(t) \rangle^{\text{exact}}_{\rm eq} \!=\!\int_{-\infty}^{\infty} \!\frac{ d\Omega}{2\pi} \!e^{i\Omega \tau} \!\frac{  J(\Omega) n_{\eta}(\Omega)}{\big(\Omega-\omega_0-\Sigma(\Omega)\big)^2 + \frac{J^2(\Omega)}{4}}.
\end{equation}
where the frequency shift term is given by $\Sigma(\Omega)= P \big[\int_{-\infty}^{\infty} \frac{d\Omega'}{2 \pi} \frac{J(\Omega')}{\Omega-\Omega'}\big]$.
Let us now consider the fermionic case and  fix $J(\Omega)$ as a constant i.e., $J(\Omega)=\Gamma$. 
This immediately gives $\Sigma(\Omega)=0$ which implies 
\begin{equation} \label{exact}
		\langle a^\dagger(t+\tau)a(t) \rangle^{\text{exact}}_{\rm eq} \!=\int_{-\infty}^{\infty}  \frac{ d\Omega}{2\pi} \,e^{i\Omega \tau} \, \frac{  \Gamma  \, n_{+1}(\Omega)}{(\Omega-\omega_0)^2 + \frac{\Gamma^2}{4}}.
\end{equation}
Recall that $n_{+1}$ corresponds to the Fermi-Dirac distribution function.
In this case, remarkably, using MQRT the correlator 
$\langle a^\dagger(t+\tau)a(t) \rangle^{\text{MQRT}}_{\rm eq}$ in \eqref{eqada} matches with the exact answer in \eqref{exact}.
 Thus the derived MQRT displays a significant advantage over the SQRT. We next consider the case when $J(\Omega)$ is frequency dependent. In this case, the MQRT result does not match the exact answer. However, the answer predicted by MQRT always remains closer to the exact result compared to the SQRT. In Fig.~\ref{fig:0} we make a comparison plot of the answers obtained following SQRT and MQRT with the exact result by defining a quantity \cite{Thingna-cano} 
\begin{equation}
D= \frac{1}{\tau_f}\int_{0}^{\tau_f} d\tau \Bigg[\Big|C_{R}(\tau)^{\text{exact}} \!-\!C_{R}(\tau)\Big| + \Big|C_{I}(\tau)^{\text{exact}} \!-\!C_{I}(\tau)\Big|\Bigg]
\label{diff}
\end{equation}
where we denote $C(\tau)= \langle a^\dagger(t+\tau)a(t)\rangle_{\rm eq}$ and $C_R, C_I$ are the real and imaginary components of the correlation function. We set the value of the upper integration limit $\tau_f$ such that the contribution of the integrand to the quantity $D$ is most significant.
It is clear from the density plots for both the dissipative boson and fermion models that the results predicated following the MQRT ((a) and (c) of Fig.~\ref{fig:0}) remain closer than the ones obtained following the SQRT ((b) and (d) of Fig.~\ref{fig:0}).


\section{Summary}
\label{Summary}
One important consequence of thermalization in the context of open quantum systems is that the multi-time correlation functions should satisfy the KMS condition.
In this work, we investigated the consistency of two-time correlation functions from the perspective of the KMS condition. We show that the correlation functions obtained following the well-known QRT fails to preserve the KMS condition at the finite order of system-bath coupling. We then point out a weaker Markov condition that helps us to obtain a modified version of QRT using which we obtain two-point function which in the long-time limit respects the KMS condition to the non-zero order of the system-bath coupling. We further showed that as compared to SQRT, the derived MQRT provides better results when compared to the exact results. 
It is worth mentioning that the result obtained by solving MQRT does not always reproduce ${\mathcal O}(\lambda^2)$ answer exactly, as also shown in Fig.~\ref{fig:0}. 
It would be interesting to explore if one can systematically improve the result obtained from MQRT to account for this discrepancy. To do this it will be useful to follow the approach recently implemented in Ref.~\onlinecite{Thingna-cano}. 


Even though in this work, we primarily focused on the two-point correlation function, our approach to derive MQRT can be suitably extended for multi-point correlation functions and KMS conditions can be systematically preserved (please see Appendix \ref{app-3}). Along these lines, it would be interesting to investigate OTOC under weak Markov approximation.

\section*{Acknowledgements} 
The work of SJ is supported by Ramanujan Fellowship. SK acknowledges the CSIR fellowship with Grant Number 09/0936(11643)/2021-EMR-I.   BKA acknowledges the MATRICS grant MTR/2020/000472 from SERB, Government of India. BKA also thanks the Shastri Indo-Canadian
Institute for providing financial support for this research work in the form of a Shastri Institutional Collaborative Research Grant (SICRG). The authors would also like to thank the people of India for their steady support in basic research. 
 
 \bibliographystyle{apsrev4-2}
 \bibliography{references.bib}
\appendix

\section{Detailed derivation of the modified QRT (MQRT) }\label{app-1}
In this section, we provide the detailed derivation of the modified QRT (MQRT). Recall that our derivation holds for the arbitrary system Hamiltonian which is interacting with a bath via a generic system operator $S$. The total Hamiltonian for our setup is given by,
\begin{equation} \label{eq1nma}
		\begin{split}
			H & = H_{S}+H_{R}+ H_{S  R} \\
			& = H_{S}+\sum_{k} \Omega_{k} b_{k}^\dagger b_{k}+ \lambda \sum_{k} \alpha_{k}( S b_{k}^\dagger+ S^\dagger   b_{k})\;,
		\end{split}
\end{equation}
where $b_{k} $ and $b_{k}^\dagger $ represents  the bosonic or fermionic annihilation and creation operator respectively. The parameter $\lambda$ is introduced here to keep track of the order of the perturbation. 
We want to calculate two-point function of the form- 
$ C(t+\tau,t)=\langle A_{\mu}(t+\tau)O(t) \rangle$, where $ \langle A_{\mu}(t)\rangle$ satisfies the following relation
\begin{align}
		\frac{d}{dt} \langle A_{\mu}(t)\rangle =\sum_{\kappa} M_{\mu \kappa}\langle A_{\kappa}(t)\rangle.
\end{align}
To calculate $ C(t+\tau,t) $, we are going to first express it in terms of one-point reduced operators.
In Ref.~\onlinecite{karve2020heisenberg}, it was shown that the two-point reduced operator upto $\lambda^2$ order can be written  as
\begin{widetext}
\begin{equation}
\label{two-reduced}
	[A_{\mu  }(t+\tau)O(t)]_{S}=A_{\mu S}(t+\tau)O_{S}(t)+ I[A_{\mu S}(t+\tau),O_{S}(t)]\;,
\end{equation}
where $A_{\mu S} $ and $O_{S} $ are the reduced one-point operators whose dynamics is determined by the adjoint Lindblad equation.  $I[A_{\mu S}(t+\tau),O_{ S }(t)]$ is called the irreducible term which can be written in terms of one-point reduced operator as
\begin{align} \label{eqgI}
	I[A_{\mu S}(t+\tau),O_{ S }(t)]  \!&=\lambda^2\int_{0}^{t+\tau} \!d\tau_{1}\! \int_{0}^{t}\! d\tau_{2}\; \beta(\tau_{2}\!-\!\tau_{1}) \Big[ A_{\mu S}(t+\tau),
	\Tilde{S}(-\tau_{1})\Big]\Big [ \Tilde{S}^\dagger(-\tau_{2}), O_{ S}(t)
	\Big]\nonumber\\
	&+\lambda^2\int_{0}^{t+\tau} \!d\tau_{1}\! \int_{0}^{t}\! d\tau_{2}\; \alpha(\tau_{2}\!-\!\tau_{1})\Big[ A_{\mu S}(t+\tau),
	\Tilde{S}^\dagger(-\tau_{1})\Big]\Big [ \Tilde{S}(-\tau_{2}), O_{  S}(t)
	\Big]\;,
\end{align}
where $\Tilde{S}(t)$ is the interaction picture operator which is defined as $\Tilde{ S}(t)= e^{-iH_{S}t} S e^{iH_{S}t}$ and it can be decomposed as $ \Tilde{S}(t) =\sum_{m} S_{m} e^{i\omega'_{m} t}$, whereas $\alpha(\tau)$, $\beta(\tau)$  are related to the bath correlation function or more explicitly
\begin{align}
	\label{correlation1}
	&\alpha(\tau)=\sum_{k}  |\alpha_{k}|^2 \, {\rm Tr}_{R} \Big[b_{  k}\Tilde{b}^\dagger_{k}(-\tau)\rho_{R}\Big] =\sum_{k}  |\alpha_{k}|^2 \big(1-\eta \;n_{\eta}(\Omega_{k})\big)e^{i\Omega_{k}\tau},\nonumber\\
	& \beta(\tau)=\sum_{k}  |\alpha_{k}|^2 \, {\rm Tr}_{R} \Big[b^\dagger_{  k}\Tilde{b}_{k}(-\tau)\rho_{R}\Big] =\sum_{k}  |\alpha_{k}|^2  n_{\eta}(\Omega_{k}) e^{-i\Omega_{k}\tau},
\end{align}
 here $n_{\eta}(\Omega)$ represents the Bose or Fermi distribution functions i.e., $n_{\eta}(\Omega)=[e^{\beta \Omega}+\eta ]^{-1}$ with  $\eta=+1$ and $\eta=-1$ are for fermions and bosons, respectively.
By taking derivative of the the Eq.\eqref{two-reduced} with respect to $\tau$ and using Eq.~\eqref{eqgI}, we can  show that the reduced two-point operator, $	[A_{\mu  }(t+\tau)O(t)]_{S}$,  obeys the following equation
\textcolor{black}{\begin{align} \label{eqqrtcc-app}
		\frac{d}{d\tau}[A_{\mu  }(t+\tau)O(t)]_{S} \!=&\sum_{\kappa}M_{\mu\kappa}[A_{\kappa }(t+\tau)O(t)]_{S} +C_{1}(t, \tau)+C_{2}(t, \tau).
\end{align}}
Note that, so-far we have only taken the Born approximation i.e., we have considered terms up to $\lambda^2$ order but have not incorporated the Markovian approximation. In the above equation $C_{1}(t, \tau)$  and  $C_{2}(t, \tau)$ are the non-homogeneous terms which are given by the following equation
\begin{align}\label{eqc1a}
	&C_{1}(t, \tau)= \lambda^2 \int_{0}^{t}\! d\tau_{2}\; \beta(\tau_{2}\!-\!t\!-\!\tau) \Big[ A_{\mu S}(t+\tau),
	\Tilde{S}(-t\!-\!\tau)\Big]\Big [ \Tilde{S}^\dagger(-\tau_{2}), O_{S}(t)\Big],
 \nonumber\\
&	C_{2}(t, \tau)= \lambda^2 \int_{0}^{t}\! d\tau_{2}\; \alpha(\tau_{2}\!-\!t\!-\!\tau_{1})\Big[ A_{\mu S}(t+\tau),
	\Tilde{S}^\dagger(-t\!-\!\tau)\Big]\Big [ \Tilde{S}(-\tau_{2}), O_{S}(t)
	\Big].
\end{align}
In the absence of the bath, the dynamics of the system operators are unitary and are governed by the system's bare Hamiltonian $H_{S}$.
More explicitly, 
$ A^{0}_{\mu S}(t)= e^{iH_{S}t}  A_{\mu S}(0)e^{-iH_{S}t}
=\sum_{j}e^{-i\omega_{j}t} A_{\mu j S}(0)$
and $ O^{0}_{ S}(t)= e^{iH_{S}t} \; O_{ S}(0)e^{-iH_{S}t}
=\sum_{j}e^{-i\tilde{\omega}_{l}t} O_{l S}(0)$. This ensures us to write down the following equations
\begin{align}
   & A_{\mu S}(t)=\sum_{j}  A_{\mu j  S}(t-t') e^{-i\omega_{j}t'}+O(\lambda^2),\nonumber\\
    & O_{S}(t)=\sum_{l}  O_{l S}(t-t') e^{-i\tilde{\omega}_{l}t'}+O(\lambda^2).
\end{align}
The $C_{1}(t,\tau)$ term defined in Eq.\eqref{eqc1a} is already $\lambda^2 $ order so if we substitute the above equation in Eq.\eqref{eqc1a}, the  $C_{1}(t,\tau)$ will be correct upto $\lambda^2$ order  and it is given by
\begin{align}
	C_{1}(t, \tau)= \lambda^2  \sum_{j,l,m,n} \int_{0}^{t}\! d\tau_{2}\; \beta(\tau_{2}\!-\!t\!-\!\tau) \Big[ A_{\mu j S}(0),
	S_{m}\Big]\Big [ S^\dagger_{n}, O_{l S}(t-\tau_{2})\Big] e^{-i(\omega_{j}+\omega'_{m})(t+\tau)} e^{-i(\tilde{\omega}_{l}-\omega'_{n})\tau_{2}}+O(\lambda^4).
\end{align}
Let's define, $\tau'_{2}=t-\tau_{2}$,
 from the definition of $\tau'_{2} $ it is clear that the limit of $ \tau'_{2}$ will be from $t$ to $0$. In terms of this new variable $ $ the above equation becomes
\begin{align}
	C_{1}(t, \tau)= \lambda^2 \sum_{j,l,m,n}  \int_{0}^{t}\! d\tau'_{2}\; \beta(-\tau\!-\!\tau'_{2}) \Big[ A_{\mu j S}(0),
	S_{m}\Big]\Big [ S^\dagger_{n}, O_{l S}(\tau'_{2})\Big] e^{-i(\omega_{j}+\omega'_{m})(t+\tau)} e^{-i(\tilde{\omega}_{l}-\omega'_{n})(t-\tau'_{2})}+O(\lambda^4).
\end{align}
Similarly, by following the identical steps we can simplify the $ C_{2}(t, \tau)$ term. Finally, if we substitute $\alpha $ and $\beta$ which are defined in Eq.\eqref{correlation1} and neglect terms higher than $\lambda^2 $ order, we get
\begin{align}
	&C_{1}(t, \tau)= \lambda^2 \sum_{j,l,m,n}  e^{-i(\omega_{j}+\omega'_{m})(t+\tau)} e^{-i(\tilde{\omega}_{l}-\omega'_{n})t}\int_{0}^{t}\! d\tau'_{2}\;  e^{i(\tilde{\omega}_{l}-\omega'_{n})\tau'_{2}}  \Big[ A_{\mu j S}(0),
	S_{m}\Big]\Big [ S^\dagger_{n}, O_{l S}(\tau'_{2})\Big] \int \frac{ d\Omega}{2\pi}  F_{\eta}(\Omega)\;e^{i\Omega (\tau+\tau'_{2})},\nonumber\\
&	C_{2}(t, \tau)= \lambda^2 \sum_{j,l,m,n}  e^{-i(\omega_{j}+\omega'_{m})(t+\tau)} e^{-i(\tilde{\omega}_{l}-\omega'_{n})t}\int_{0}^{t}\! d\tau'_{2}\;  e^{i(\tilde{\omega}_{l}-\omega'_{n})\tau'_{2}}  \Big[ A_{\mu j S}(0),
	S_{m}\Big]\Big [ S^\dagger_{n}, O_{l S}(\tau'_{2})\Big]  \int \frac{ d\Omega}{2\pi} \Big(J(\Omega)-\eta F_{\eta}(\Omega)\Big)\;e^{-i\Omega (\tau+\tau'_{2})},
\end{align}
where $J(\Omega)$ is the spectral density function of the bath which is defined as $J(\Omega)= 2\pi \sum_{k} |\alpha_{k}|^2 \delta(\Omega-\Omega_{k})$ and $F_{\eta}(\Omega)=J(\Omega) n_{\eta}(\Omega)$.
If we take the following Markovian approximation that $t+\tau \gg \tau_{B}$, where $\tau_{B}$ is the bath characteristic time scale then the upper limit of the $\tau'_{2}$ integration in $C_{1} $ and $ C_{2}$ can be extended from $t$ to $\infty$. Thus we derive the modified QRT or MQRT based on a weak Markovian approximation.   However, to get the standard QRT we need to assume a stronger Markovian approximation which is $\tau \gg \tau_{B}$. Under this  approximation both the  $C_{1} $ and $ C_{2}$  terms vanish since the $d\Omega$ integration gives zero contribution. 
\end{widetext}

\section{MQRT and KMS relation for dissipative Spin-Boson model}
\label{app-2}
In this appendix, we apply the MQRT to calculate the correlation function for another paradigmatic model, namely the dissipative spin-boson model.
\textcolor{black}{The total Hamiltonian for the dissipative Spin-Boson model is given as}
\begin{equation} \label{eq1nmea}
			H  = \frac{\omega_{0}}{2}\sigma_{z}+\sum_{k} \Omega_{k} b_{k}^\dagger b_{k}+ \sum_{k} \alpha_{k}( \sigma_{-} b_{k}^\dagger+
   \sigma_{+} b_{k}),
\end{equation}
where $b_{k} $($b^\dagger_{k} $) represents the bosonic annihilation (creation) operator and $\sigma_{-}$ ($\sigma_{+}$) is the lowering (raising) operator of the spin-half system.
For this model, using adjoint Lindblad equation \eqref{ad-QME}, we can show that $\sigma_{- S}$ and  $\sigma_{+ S}$ satisfy the following equation
\begin{align}\label{eqspco}
	& \frac{d}{dt} \sigma_{+ S}(t)=-\tilde{G} \sigma_{+ S}(t)\nonumber\\
	&     \frac{d}{dt} \sigma_{- S}(t)=-\tilde{G}^{*} \sigma_{- S}(t)\;,
\end{align}
where 
\begin{equation}
    \tilde{G}= \Big[-i\omega'_{0} + \big(\frac{\gamma}{2}+\gamma \, n(\omega_{0})\big)\Big].
\end{equation}
Here, $\omega_0'=\omega_{0}+P \big[\int \frac{d\Omega}{2\pi} \frac{J(\Omega)}{\omega_{0}-\Omega}\big]+P \big[\int \frac{d\Omega}{\pi} \frac{F(\Omega)}{\omega_{0}-\Omega}\big]$ is the new modified frequency. Recall that $J(\Omega)$ is the spectral density of the bath and $F(\Omega)= J(\Omega) n(\Omega)$ with $n(\Omega)$ is the Bose-Einstein distribution function. We note $\gamma=J(\omega_{0})$.
We are interested in computing the correlation $\langle \sigma_{+ }(t+\tau)\sigma_{- }(t)\rangle$ correlation function.
Using the MQRT expressed in Eq.~\ref{eqqrtcc}, we can show that the reduced two-point operator, $[\sigma_{+ }(t+\tau)\sigma_{- }(t)]_{S} $, 
 satisfies the following equation
\begin{align} \label{eqsb2pt}
	\frac{d}{d\tau}[\sigma_{+ }(t+\tau)\sigma_{- }(t)]_{S} \!=&-\tilde{G}[\sigma_{+ }(t+\tau)\sigma_{- }(t)]_{S} +C_{1}(\tau),
\end{align}
where the first non-homogeneous part of Eq.~\ref{eqqrtcc} i.e., $C_{1}(t,\tau)$ is given by
\begin{equation}
    C_{1}(\tau)=  \int_{-\infty}^{\infty} \frac{ d\Omega}{2\pi}  \;e^{i\Omega \tau} \frac{  F(\Omega)}{\tilde{G}^{*}-i\Omega}
\end{equation}
which is a function of $\tau$ only as all $t$ dependent terms disappear. 
Also the second non-homogeneous part of Eq. \eqref{eqqrtcc} i.e., $C_{2}(t,\tau)$ vanishes in this case because 
\begin{equation}
  C_{2}(t,\tau) \propto [\sigma_{+}, \sigma_{+}][\sigma_{-}, \sigma_{-}]=0.
\end{equation}
Using the above equation we find the two-point function at the equilibrium which is given by
\begin{eqnarray}
\label{eqada1a}
	&&\langle \sigma_{+ }(t+\tau)\sigma_{- }(t)\rangle^{\text{MQRT}}_{\rm eq}= \nonumber \\
 && \quad \quad \quad \quad \int_{-\infty}^{\infty} \frac{ d\Omega}{2\pi} e^{i\Omega \tau}\; \bigg[\frac{J(\Omega)n(\Omega)}{(\tilde{G}+i\Omega) (\tilde{G}^{*}-i\Omega)}\bigg].
\end{eqnarray}
Similarly, we obtain the other correlator, $\langle\sigma_{- }(t)\sigma_{+}(t+\tau)\rangle$, at the equilibrium using the MQRT as,
\begin{eqnarray}
    \label{eqaad1a}
	&&\langle \sigma_{- }(t)\sigma_{+ }(t+\tau)\rangle^{\text{MQRT}}_{\rm eq} = \nonumber \\
&& \quad \quad\quad \quad \quad \int_{-\infty}^{\infty} \frac{ d\Omega}{2\pi} e^{i\Omega \tau}\; \bigg[\frac{J(\Omega)(n(\Omega)+1)}{(\tilde{G}+i\Omega) (\tilde{G}^{*}-i\Omega)}  \bigg].\,\,\,\,\,\,
\end{eqnarray}
Using Eq.\eqref{eqada1a} and Eq.\eqref{eqaad1a}, it is clear that  $ \langle \sigma_{+ }(t+\tau)\sigma_{- }(t)\rangle$ satisfies
the KMS condition up to  the second order in system-bath coupling. For comparison, the SQRT predicts the following results for the above correlators
\begin{eqnarray}
	&&\langle \sigma_{- }(t)\sigma_{+ }(t+\tau)\rangle^{\text{SQRT}}_{\rm eq} = \nonumber \\
&& \quad \quad \quad e^{-\tilde{G}\tau} \int_{-\infty}^{\infty} \frac{ d\Omega}{2\pi} \; \bigg[\frac{J(\Omega)(n(\Omega)+1)}{(\tilde{G}+i\Omega) (\tilde{G}^{*}-i\Omega)}  \bigg].
\end{eqnarray}
which clearly does not respect the KMS to the non-zero order of the system-bath coupling. Similarly, for the other correlator we receive the following SQRT,  
\begin{eqnarray}
	&&\langle \sigma_{+}(t+\tau)\sigma_{-}(t)\rangle^{\text{SQRT}}_{\rm eq} = \nonumber \\
&& \quad \quad e^{-\tilde{G}\tau} \int_{-\infty}^{\infty} \frac{ d\Omega}{2\pi} \; \bigg[\frac{J(\Omega) n(\Omega)}{(\tilde{G}+i\Omega) (\tilde{G}^{*}-i\Omega)}  \bigg].
\end{eqnarray}

\section{MQRT and KMS for three-point correlation function}
\label{app-3}
In this appendix, we extend our analysis to three-point correlators.  So far we have analyzed the proper way of considering the Markovian limit (weak Markov)  for two-point correlation functions such that the KMS condition is respected in the long-time limit. However, a natural  question that immediately arises is how to understand the weak Markov limit for higher-point function as well such that the corresponding KMS conditions are respected to the non-zero order of the system-bath coupling. To investigate this question,   
we provide an example by calculating the three-point function for the non-interacting bosonic model in Eq.~\ref{CL}. It is easy to generalize the result for the fermionic case as well. In particular, we calculate the following three-point function involving the bosonic operators $ \langle a^\dagger(t+\tau_{1}+\tau_{2})a(t+\tau_{2}) N(t)\rangle$ with $N(t)=a^{\dagger}(t) a(t)$. We can show that for the three-point reduced operator of the kind $[ a^\dagger(t+\tau_{1}+\tau_{2})a(t+\tau_{2}) N(t)]_{S}$ the weak Markovian approximation implies setting $t + \tau_1 + \tau_2 \gg \tau_B$ which satisfies the following equation
\begin{widetext}
    \begin{align} 
		\frac{d}{d\tau_{1}}\big[a^\dagger(t+\tau_{1}+\tau_{2})a(t+\tau_{2}) N(t)\big]_{S}=-G \big[a^\dagger(t+\tau_{1}+\tau_{2})a(t+\tau_{2}) N(t)\big]_{S}+	C_{1}(\tau_{1},\tau_2, t)+C_{2}(\tau_{1},\tau_{2},t).
\end{align}
\end{widetext}
where once again two non-homogeneous terms $C_1, C_2$ appear as a correction to the SQRT. The solution of the above differential equation will depend on the two-point reduced operator evaluated at $\tau_1=0$ and therefore of the form $[N(t+\tau_{2})N(t)]_{S}$. Using the MQRT for the two-point function, we can easily calculate $[N(t+\tau_{2})N(t)]_{S}$. If we consider the equilibrium limit i.e., $t \to \infty$ then the three-point function at equilibrium reduces to
\begin{widetext}
\begin{align} 
	\langle a^\dagger(t+\tau_{1}+\tau_{2})a(t+\tau_{2}) N(t) \rangle^{\text{MQRT}}_{\rm eq} \!&=\int_{-\infty}^{\infty}\frac{ d\Omega}{2\pi}  \; \frac{J(\Omega)}{(G^*-i\Omega) (G+i\Omega)}
	\Big[n(\omega_{0})e^{i\Omega 
 \tau_{1}}n(\Omega)+ n(\omega_{0})e^{i\omega_{0} (\tau_{1}+\tau_{2})} e^{-i\Omega \tau_{2}}(1+n(\Omega)) \nonumber \\
 & \quad \quad \quad \quad + (1+n(\omega_{0}))e^{-i\omega_{0} \tau_{2}} e^{i\Omega (\tau_{1}+\tau_{2})}n(\Omega) \Big].
\end{align}
where $n(\Omega)$ is the Bose distribution function. 
Similarly, one can derive the following three-point function  
$ \langle  N(t)a^\dagger(t+\tau_{1}+\tau_{2})a(t+\tau_{2})\rangle$ and is given by
\begin{align} \label{eq-3point}
		&	\langle  N(t)a^\dagger(t+\tau_{1}+\tau_{2})a(t+\tau_{2})\rangle^{\text{MQRT}}_{\rm eq} \!=\int_{-\infty}^{\infty} \frac{ d\Omega}{2\pi}  \; \frac{J(\Omega)}{(G^*-i\Omega) (G+i\Omega)}\Big[  n(\omega_{0})e^{i\Omega \tau_{1}}n(\Omega)+ (1+ n(\omega_{0}))e^{i\omega_{0} (\tau_{1}+\tau_{2})} e^{-i\Omega \tau_{2}}n(\Omega)\nonumber\\
		&\quad \quad \quad \quad + n(\omega_{0})e^{-i\omega_{0} \tau_{2}} e^{i\Omega (\tau_{1}+\tau_{2})}(1+n(\Omega)) \Big].
\end{align}
Interestingly, the validity of the KMS condition involving these two correlation functions can be easily checked and in this case, there are four such relations 
\begin{align}
&\langle a^\dagger(t+\tau_{1}+\tau_{2})a(t+\tau_{2}) N(t) \rangle^{\text{MQRT}}_{\rm eq} =\langle N(t-i\beta) a^\dagger(t+\tau_{1}+\tau_{2})a(t+\tau_{2})  \rangle^{\text{MQRT}}_{\rm eq}\nonumber\\
&\langle a^\dagger(t+\tau_{1}+\tau_{2})a(t+\tau_{2}-i\beta) N(t) \rangle^{\text{MQRT}}_{\rm eq} =\langle N(t-i\beta) a^\dagger(t+\tau_{1}+\tau_{2})a(t+\tau_{2}-i\beta)  \rangle^{\text{MQRT}}_{\rm eq}\nonumber\\
&\langle a^\dagger(t+\tau_{1}+\tau_{2}-i\beta)a(t+\tau_{2}) N(t) \rangle^{\text{MQRT}}_{\rm eq} =\langle N(t-i\beta) a^\dagger(t+\tau_{1}+\tau_{2}-i\beta)a(t+\tau_{2})  \rangle^{\text{MQRT}}_{\rm eq}\nonumber\\
&\langle a^\dagger(t+\tau_{1}+\tau_{2}-i\beta)a(t+\tau_{2}-i\beta) N(t) \rangle^{\text{MQRT}}_{\rm eq} =\langle N(t-i\beta) a^\dagger(t+\tau_{1}+\tau_{2}-i\beta)a(t+\tau_{2}-i\beta)  \rangle^{\text{MQRT}}_{\rm eq}
\end{align}

\section{Calculation of $\langle a^\dagger(t)a(t)\rangle$ in the long-time limit for dissipative boson/fermion model}
\label{app-4}
In this appendix, we provide the details of the calculation for $\langle a^\dagger(t)a(t)\rangle$ in the long-time limit for the dissipative boson/fermion model as introduced in Eq.~\ref{CL}. To calculate the correlator that is correct upto the second-order of the system-bath coupling,  We write the two-point reduced operator $[a^\dagger(t)a(t)]_{S}$ as \cite{karve2020heisenberg}
  \begin{equation}
[a^\dagger(t)a(t)]_{S}=a^\dagger_{S}(t)a_{S}(t)+
 I[a^\dagger_{S}(t),a_{S}(t)],
  \end{equation}
where the second term represents the irreducible part $I[a^\dagger_{S}(t),a_{S}(t)]$ which to the second-order of system-bath coupling i.e., upto $O(\lambda^2)$ is given by
 \begin{align}\label{eq}
 I[a^\dagger_{S}(t),a_{S}(t)] =
 \int_{0}^{t} \!d\tau_{1}\! \int_{0}^{t}\! d\tau_{2}\; \beta(\tau_{2}\!-\!\tau_{1}) \Big[ a^\dagger_{S}(t-\tau_{1}),
	a \Big]\Big [ a^\dagger, a_{ S}(t-\tau_{2})
	\Big]+O(\lambda^4)
\end{align}
where recall that 
$\beta(\tau)=\sum_{k}  |\alpha_{k}|^2 \, {\rm Tr}_{R} \Big[b^\dagger_{  k}\Tilde{b}_{k}(-\tau)\rho_{R}\Big] =\sum_{k}  |\alpha_{k}|^2 n_{\eta}(\Omega_{k}) e^{-i\Omega_{k}\tau}$ is the bath-correlation function. We next define, $\tau'_{1}=t-\tau_{1}$ and $\tau'_{2}=t-\tau_{2}$, and then the above equation reduces to
 \begin{align}
 I[a^\dagger_{S}(t),a_{S}(t)]=\int_{0}^{t} \!d\tau'_{1}\! \int_{0}^{t}\! d\tau'_{2}\; \beta(\tau'_{1}\!-\!\tau'_{2}) \Big[ a^\dagger_{S}(\tau'_{1}),
	a \Big]\Big [ a^\dagger, a_{ S}(\tau'_{2})
	\Big]
\end{align}
Now by substituting the expression for the bath correlation function $\beta(\tau)$ and performing the $\tau'_{1}$, $\tau'_{2}$ integration we obtain,
\begin{align}\label{eqi1}
I[a^\dagger_{S}(t),a_{S}(t)]=\int_{-\infty}^{\infty} \frac{ d\Omega}{2\pi}  \bigg[\frac{1-e^{-(G+i\Omega)t}}{(G+i\Omega)} F_{\eta}(\Omega)\frac{1-e^{-(G^{*}-i\Omega)t}}{(G^{*}-i\Omega)}  \bigg].
\end{align}
Finally, the two-point reduced operator $[a^\dagger (t)a(t)]_{S}$ at the long-time limit is given by 
\begin{align}\label{eqiap}
\lim_{t \to \infty} [a^\dagger(t) a(t)]_{S}= \int_{-\infty}^{\infty} \frac{ d\Omega}{2\pi} \bigg[\frac{F_{\eta}(\Omega)}{(G+i\Omega) (G^{*}-i\Omega)}  \bigg].
\end{align}
This expression is used in obtaining the long-time limit result in Eq.~\ref{eqada}.
\end{widetext}

\newpage

\end{document}